# Titania Prepared by Ball Milling: Its Characterization and Application as Liquefied Petroleum Gas Sensor

B.C. Yadav[1,2]*, Tripti Shukla[1], Satyendra Singh[1] and T.P. Yadav[3]

[1]Nanomaterials and Sensors Research Laboratory, Department of Physics,
University of Lucknow, Lucknow-226007, U.P., India

[2]Department of Applied Physics, School of Physical Sciences,
Babasaheb Bhimrao Ambedkar University, Lucknow-226025

[3]Department of Physics, Banaras Hindu University, Varanasi, U.P., India

*Corresponding author: Mobile: +919450094590
Email:balchandra_yadav@rediffmail.com

## ABSTRACT

Present paper reports the LPG sensing of $TiO_2$ obtained through ball milling. The milled powder was characterized by XRD, TEM and UV-visible spectroscopy. Further the ball milled powder was compressed in to pellet using hydraulic press. This pellet was investigated with the exposure of LPG. Variations in resistance with exposure of LPG to the sensing pellet were recorded. The sensitivity of the sensor was ~ 11 for 5 vol.% of LPG. Response and recovery times of the sensor were ~ 100 and 250 sec. The sensor was quite sensitive to LPG and results were found reproducible within ± 91%.

## 1. Introduction

Solid-state sensors are most versatile sensors, as they detect a wide variety of gases, and can be used for various applications. Different response characteristics are achieved by varying the semiconductor materials, processing techniques and sensor operating temperature. Among the unique attributes of the solid-state sensor the abilities of the sensors are to detect low ppm levels of gases as well as high combustible levels [1-2]. Solid-state oxide gas-sensing devices function by measuring the change in electrical resistance of a metal oxide pellet as a function of varying gas concentrations [3-5]. The adsorption (physisorption or chemisorption) of the gas species on the surface of the metal oxide results in the change in surface coverage which is responsible for the change in the surface electrical resistance. Taking into consideration that the sensing phenomena mainly takes place on the material surface, the surface morphology has an essential role on the sensitivity of solid state sensor. In the last few years, the nanograined materials offer new opportunities for enhancing the properties and performances of gas sensors. Several research reports [6–10] have confirmed the beneficial effect of nanostructure on the sensor performance.

Recently, the semiconductor gas sensors are broadly applied to the atmosphere monitoring system, the toxic or explosive gases detection system, the chemical processing facilities, the intelligent buildings with environmental control functions and so on. LPG is one of extensively used but potentially hazardous gas. Therefore, its detection is important for domestic appliances because explosion accident may be caused when it leaks out accidentally or by mistake [11-16]. $TiO_2$ is the potential material for the candidates for gas detection because of its low cost, stable phase, desirable sensitivity, high chemical stability, amenability to doping, nontoxicity and has commercially been used as lambda sensor in exhaust pipes [17-23].

In the present work we have prepared the $TiO_2$ nanoparticles using ball milling and condensed the powder in the form of a pellet. Further this pellet was investigated as LPG sensor at room temperature, which would be proven robust, cost effective and more sensitive than previous sensors [14, 24, 27].

## 2. Experimental Details

TiO$_2$ was synthesized by using high-purity (99.9%) elemental powder of Ti of ~ 0.5 mm size. This was mechanically milled in a high-energy attritor ball mill (Szegvari Attritor). The attritor has a cylindrical stainless steel tank of inner diameter 13 cm. The grinding balls made of stainless steel are of 6 mm in diameter. Ball milling was carried out for 10 hrs at 400 rpm with ball to powder ratio of 40:1 for preparing TiO$_2$ nanoparticles. The pelletization of the ball milled powder was done using hydraulic press machine (MB Instrument, Delhi) under a pressure of 616 MPa at room temperature. This pellet was put within the Ag-Pellet-Ag electrode configuration. Further it was exposed to LPG in a self-designed conventional chamber and corresponding variations in resistance with the time were recorded by using a Keithley Electrometer (Model: 6514, Germany).

## 3. Characterization Technique

### *3.1 X-Ray Diffraction*

Figure 1 shows the X-Ray Diffraction (XRD) pattern of the mechanically milled TiO$_2$ powder. The intensive peaks centered at 27°, 36° and 55° indicating the formation of TiO$_2$ in the rutile phase. From the figure it is clear that all the peaks are of the rutile phase, no other phase was identified. It can be observed that there are sharp diffraction peaks which indicate the sample has crystalline nature and no amorphous phase has been formed during present milling conditions. This result suggested that the nano-TiO$_2$ powder is composed of irregular polycrystalline.

### *3.2 Transmission Electron Microscopy*

Transmission electron microscopy (TEM) studies using a Technai 20 $G^2$ operating at 200kV were carried out in order to confirm the morphology, particle size and the structure of the phases evolved during milling. Figure 2(a) shows bright field transmission electron micrograph of 10 hrs ball milled $TiO_2$ powder. The homogeneous nano particle with spherical morphology can readily be seen in the microstructure. The $TiO_2$ particles are highly strained due to the ball milling and it can be seen in form of intergrowth in figure 2(a). The size distribution of nanoparticle has been observed between 10-20 nm. The corresponding selected area diffraction pattern in figure 2(b) show a spotty rings, it is due to the nano crystalline nature of $TiO_2$ particle.

### *3.3 UV-visible Spectroscopy*

Optical characterization of the sensing element was done by using UV-visible spectrophotometer (Varian, Carry-50Bio). Figure 3 shows UV-visible absorption spectra of titanium oxide in UV and visible range. Titanium oxide nanoparticles reveal a strong change of their optical absorption when their size is reduced. Therefore, absorption spectra of titanium oxide nanoparticles obtained in the UV-visible region show blue shift in the absorption edge at 268 nm as compared to bulk. The corresponding band gap was found 4.46 eV respectively. It is evident that titanium oxide shows significant blue shift of the absorption peak relative to the bulk absorption. This blue shift is useful for gas sensing applications.

## 4. Device Assembly

The main goal of this investigation is to develop a high sensitive LPG sensor and at the same time to analyze the effect of parameters on the gas sensing properties of the pellet. The purpose is not only to evaluate the gas sensing response, but also to reproduce as much as possible, a real environment for a working LPG sensor. The schematic diagram of experimental setup is shown in Figure 4(a). The heart part of the device is a resistance measuring pellet holder. It is well fitted in a glass chamber having gas inlet and outlet knobs for LPG. Inlet knob is associated with the concentration measuring system along with a thermocouple [24].

**Concentration Measuring System**

Figure 4(b) consists of a glass bottle containing double distilled water, which is saturated with LPG, in order to avoid the possibility of dissolution of inserted gas above this bottle, the measuring tube (pipette) is connected by vacuum seal. The cock I is connected to the LPG cylinder and cock II is connected to the inlet of the gas chamber. When the cock I is opened, the LPG from the cylinder is filled in the glass bottle and an equivalent amount of water is displaced in the measuring pipette. When the cock II is opened, a desired amount of gas e.g. 1, 2, 3 vol.% and onwards is cast out in the gas chamber.

## 5. Gas Sensing Properties

The pellet of the sensing element is subjected to exposition of LPG. Variations in resistance with the variation of the time after exposure in seconds have been recorded. The gas sensitivity is defined as

$$S = \frac{R_g}{R_a}$$

Where $R_a$ & $R_g$ stand for the resistance of the sensor in air and in the sample gas respectively [29].

Percentage sensor response of the gas sensor is defined as

$$S.R.\% = \left| \frac{Ra - Rg}{Ra} \right| \times 100$$

## 6. Results and Discussion

Prima-facie before the exposition of LPG to the sensing element, the gas chamber was allowed to evacuate at room temperature for 15-20 min and the stabilized resistance was taken as $R_a$. The variations in resistance with the time for different concentrations of LPG were observed. Figure 5 illustrates the variations in resistance of pellet with time after exposure for different vol.% of LPG at room temperature. From the figure it is clear that as time increases the resistance of the sensing pellet increases drastically in the beginning after that it becomes saturated. Finally, when we open the outlet of the chamber, the resistance approaches to their initial value of stabilized resistance in air ($R_a$) for further range of time. Curve 'a' for 1 vol.% of LPG shows constant variation in resistance with time after exposure having low sensor response. Curve 'b' for 2 vol.% of LPG exhibits improvement over the previous and the sensing element has better sensor response. Curve 'c' for 3 vol.% of LPG shows as time after exposure increases, resistance increases up to 350 sec, after that it becomes constant. Further for 4 and 5 vol.% of LPG resistance increases with time after exposure up to 350 and 400 sec respectively. Figure 6 illustrates the variations in average sensitivity of $TiO_2$ pellet for different vol.% of LPG and it is found that as the concentration of LPG (in vol.%) inside the chamber increases, average sensitivity of the sensor increases linearly. The maximum sensitivities were found 10.22 and 11 respectively for 4 and 5 vol.% of LPG. Sensor response curves are shown in Figure 7 and the maximum values of sensor responses were 1015 and 1072 for 4 and 5 vol.% of LPG. Figure 8 shows the reproducibility curve for 5 vol.% of LPG. Results were found reproducible within ± 91% accuracy. Response and recovery times were calculated as ~ 100 and 250 sec. These are important factors for gas sensor, when the sensor is exposed to and later removed from the gas environment. The response of nanostructured materials is directly related to exposed surface volume. Here the response time required for the response value to attain 90% of its maximum value is shorter (~ 100 sec). As the LPG was turned-off, the response fell rapidly, indicating the good recovery of the resistance for the material. The time taken by the sensor elements to come back once the LPG was removed is found (~ 250 sec).

The solid-state gas sensing mechanism of $TiO_2$ based LPG sensor is a surface controlled phenomenon i.e., it is based on the surface area of the pellet at which the LPG molecules adsorb and reacts with the pre adsorbed oxygen molecules. It includes the microstructure of the sensing surface and the chemical interactions occurring on the surface. The microstructure of a sensor material is important for two reasons: first, microstructure affects the distribution of adsorption sites; second, microstructure will affect electrical transport properties. The crystallite size, surface states, oxygen adsorption and the lattice defects play important role in the gas sensing. If the crystallite size is decreased, the gas sensitivity would be increased, as it was observed that the gas sensitivity was inversely proportional to the crystallite size. As $TiO_2$ is a n-type semiconductor, therefore the species who tend to trap electrons from the semiconductor are those who are easily adsorbed. Thus, adsorbed oxygen species transform at the surface of an oxide according to the general scheme $O_2, O_2^-, O^-$ or $O^{2-}$ in which they are gradually becoming richer in electrons. This may be explained by the following chemisorptions reaction:

$$O_2(gas) \leftrightarrow O_2(ads)$$
$$O_2(ads) + e^- \rightarrow O_2^-$$

The electron transfer from the conduction band to the chemisorbed oxygen results in the decrease of the electron concentration at the surface of pellet. As a consequence, there is an increase in the resistance. After some time equilibrium state is achieved between oxygen of titanium oxide and
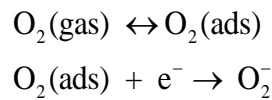

atmospheric oxygen at room temperature through chemisorptions. The stabilized resistance at present state is known as resistance in presence of air ($R_a$).

Metal oxide semiconductors are mainly used to detect small concentrations of LPG in air. The detection mechanism of LPG requires oxygen in the atmosphere and is influenced by the presence of water vapor. Some metal oxides are intrinsically n-type semiconductors, the stoichiometric excess of metal being due to oxygen vacancies. Under normal sensor operating conditions, the conductivity of the surface have revealed to be much lesser than that of the bulk. This has been attributed to the formation of surface oxygen ions that trap electrons, inducing a surface depletion layer and thus the development of Schottky barrier at inter particle contacts. This barrier exercises controlling influences over the conductivity and its height $\Phi_s$ is related to the square of the surface charge $(N_s\theta)^2$ by the expression:

$$\Phi_s = \frac{e\,(N_s\theta)^2}{2\in\in_o N_D}$$

where $N_s$ is the number of surface states per unit area, $\theta$ is the fractional occupancy and $N_D$ is the number of donor states per unit volume in the bulk. There are number of studies which have been devoted to identify the surface oxygen species. Below about 200°C, the stable surface ions are $O_2^-$. Between 200 °C and 500 °C the principal ion present is $O^-$. At temperature $\geq$ 550 °C, equilibrium of lattice oxygen with the atmosphere occurs.

In case of $TiO_2$ pellet, as the electrons entered from ionized donors through the conduction band, the charge carrier density at the interface is thereby reduced and a potential barrier to charge transport would be developed. As the surface charge grows, the adsorption of further oxygen is inhibited, the adsorption rate slows down because charge must be transferred to the adsorbate over the developing surface barrier and the coverage saturates at a quite low value. At the junction between the grains of the sensing element, the depletion layer and associated potential barrier lead to high resistance contracts, this dominates the resistance of the $TiO_2$ pellet. The resistance is then sensitive to the coverage on the surface of adsorbed oxygen ions and any factor that changes this, will change the resistance further. For instance in the presence of reactive gas a surface catalyzed combustion might occur and the surface coverage of the adsorbed oxygen ions might be decreased and the resistance would correspondingly changes. As a consequence of the reduction of potential barrier height $\Phi_s$, the depletion layer width $D_s$, the charge $Q_s$ associated with oxygen surface coverage and the donor density $N_D$ in the metal oxide are related:

$$D_s = \frac{Q_s}{eN_D}$$

However, the processes of surface catalysis and electrical response are not necessarily always connected, since for a surface reaction to give an electrical response, there must be some coupling with the charge carrier concentration [25-26].

When $TiO_2$ pellet is exposed to reducing gas like LPG, the LPG reacts with the chemisorbed oxygen. On interaction with alkanes of LPG the adsorbed oxygen is removed, forming gaseous species and water. Consequently, the resistance changes, which is due to the change in the width of depletion layer after exposure to LPG. The overall reaction of LPG with the chemisorbed oxygen may takes place as shown below:

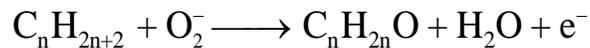

$$C_nH_{2n+2} + O_2^- \longrightarrow C_nH_{2n}O + H_2O + e^-$$

Where $C_nH_{2n+2}$ represents the various hydrocarbons. These liberated electrons recombine with the majority carriers of sensing pellet resulting in a decrease in conductivity. The formation of barrier is due to reduction in the concentration of conduction carriers and thereby, results in an increase in resistance of the sensing element with time. As the pressure of the gas inside the chamber increases, the rate of the formation of such product increases and potential barrier to charge transport becomes strong which has stopped the further formation constituting the resistance constant [27].

## 7. Conclusion

The physical synthesis of $TiO_2$ through ball milling gave uniform shape and size and has proven to be an economical and robust process. The XRD pattern clearly identifies that only tetragonal $TiO_2$ phase is present. The estimated optical band gap of $TiO_2$ was 4.46 eV which is significant for gas sensing point of view. The maximum sensitivity of $TiO_2$ was 11 for 5 vol.% of LPG. The percentage sensor response was also evaluated and its maximum value was 1072 for 5 vol.% of LPG. The sensing characteristic of $TiO_2$ pellet was 91% reproducible. The response and recovery time of sensing pellet were found ~ 100 sec and 250 sec respectively. Thus, this study demonstrates the possibility of utilizing $TiO_2$ nanomaterial prepared by ball milling as a sensing element for the detection of LPG at room temperature.

## Acknowledgement

Financial support from U.G.C. as Major Research Project F.No.36-265/2008 (SR) is highly acknowledged.

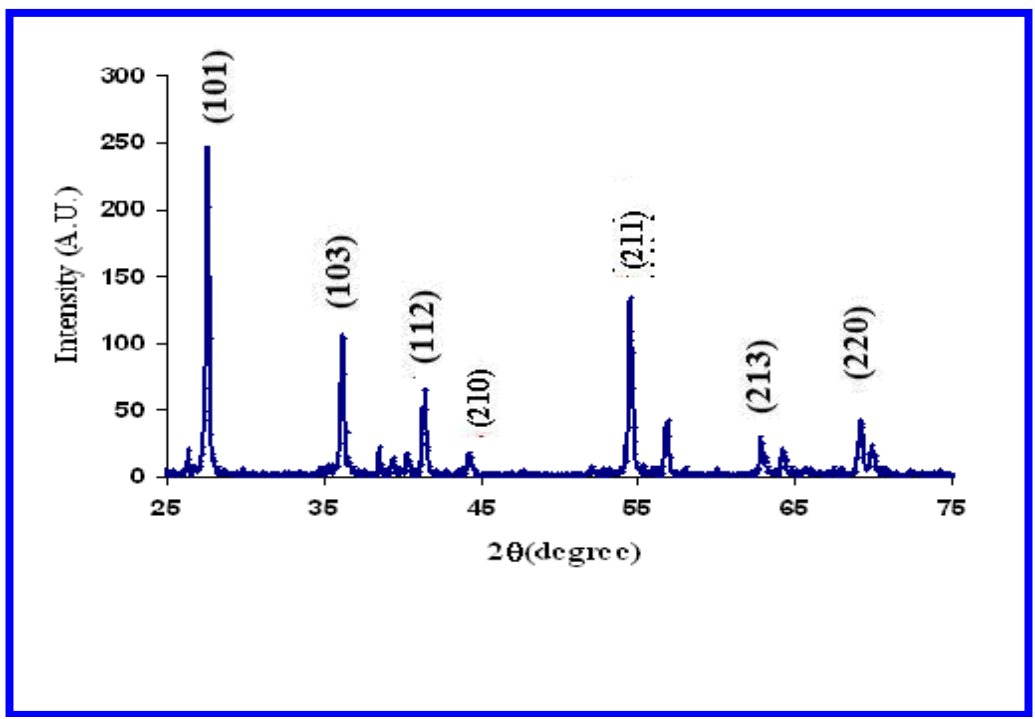

**Figure 1** X-ray Diffraction pattern of TiO$_2$ in the form of powder.

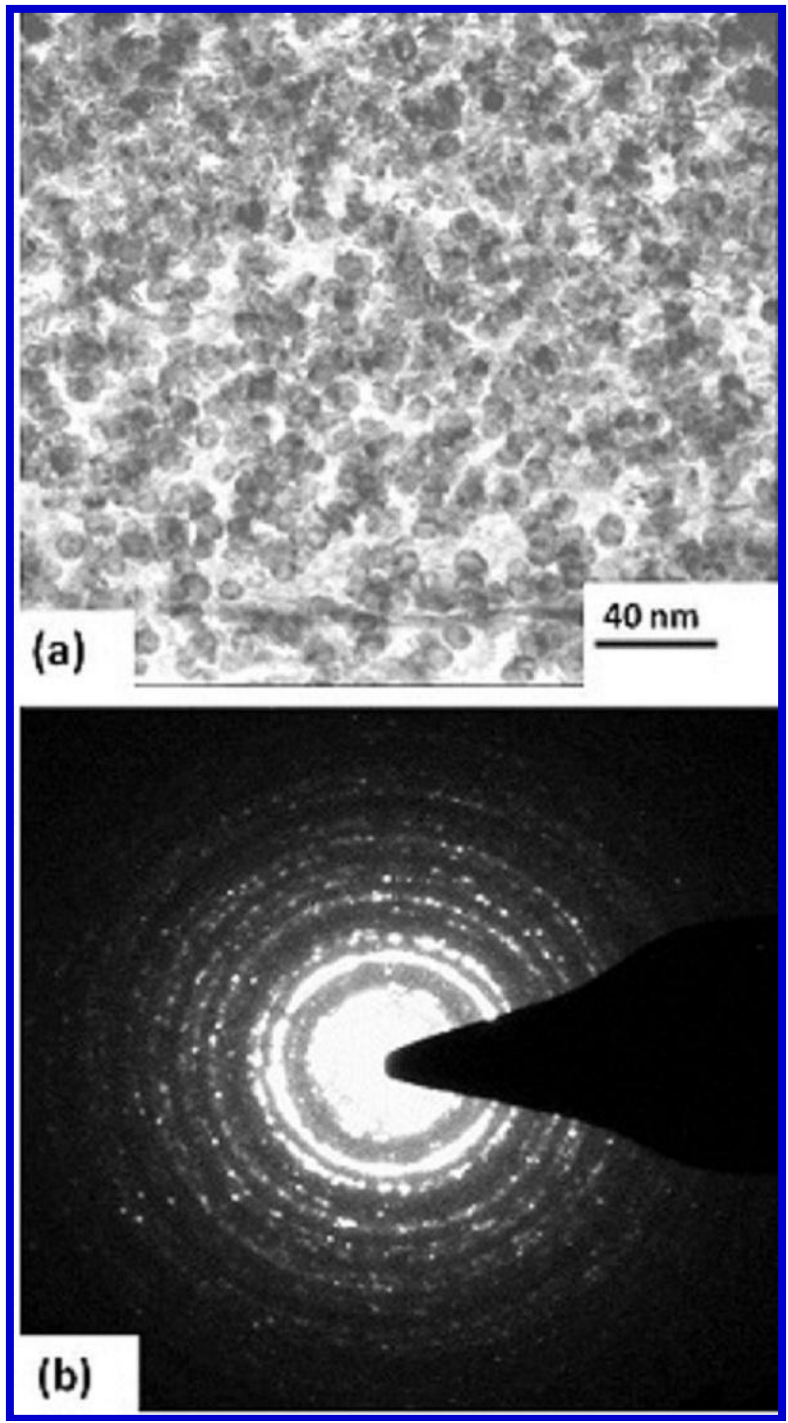

**Figure 2** TEM of $TiO_2$ in the form of powder.

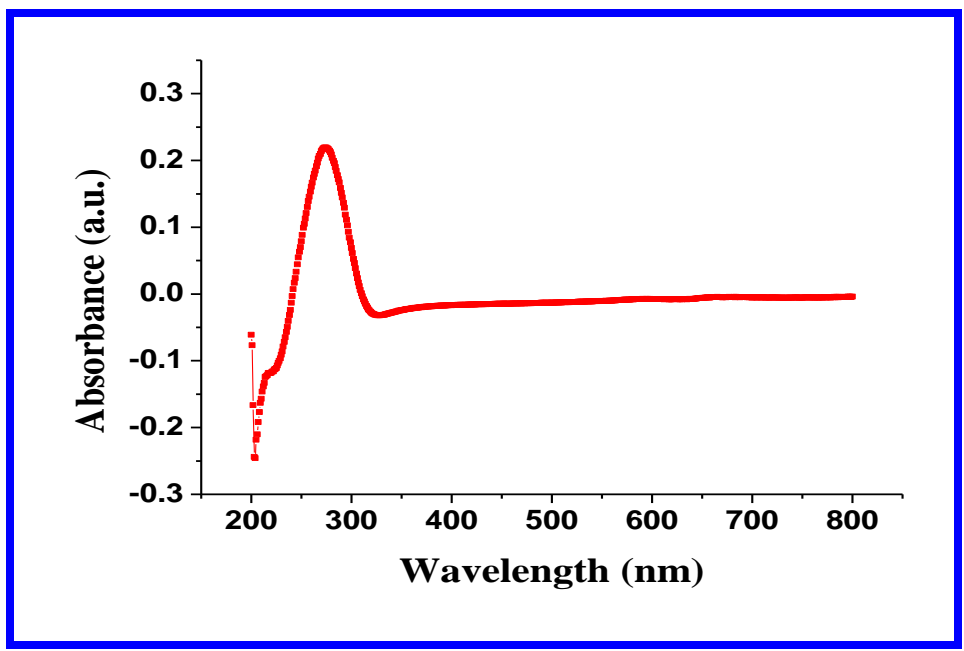

**Figure 3** UV-Visible absorption spectra of TiO$_2$ material.

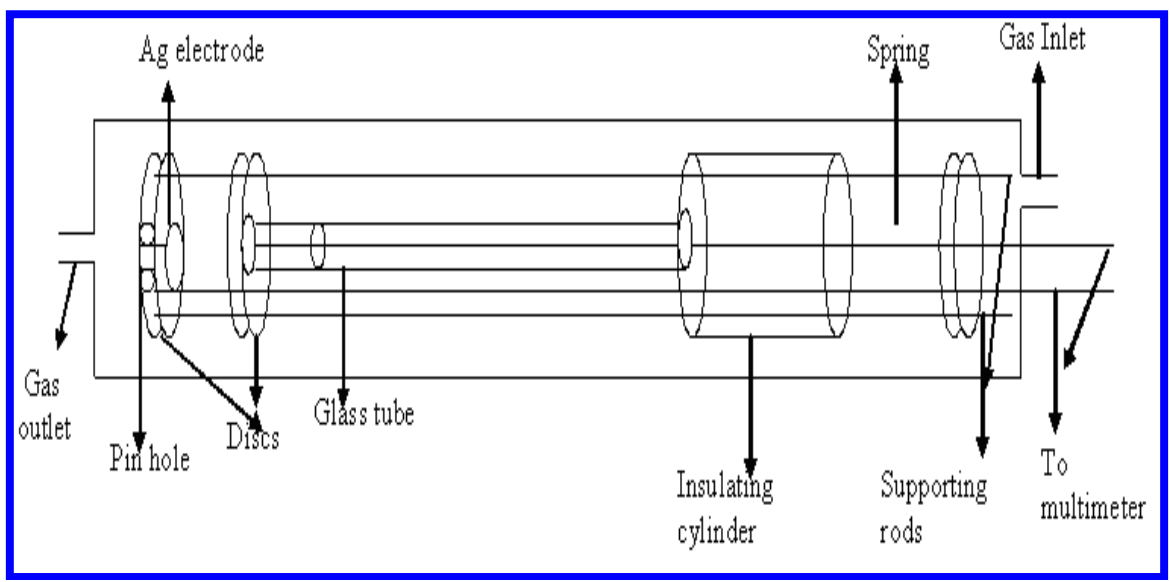

**Figure 4(a)** Device Assembly

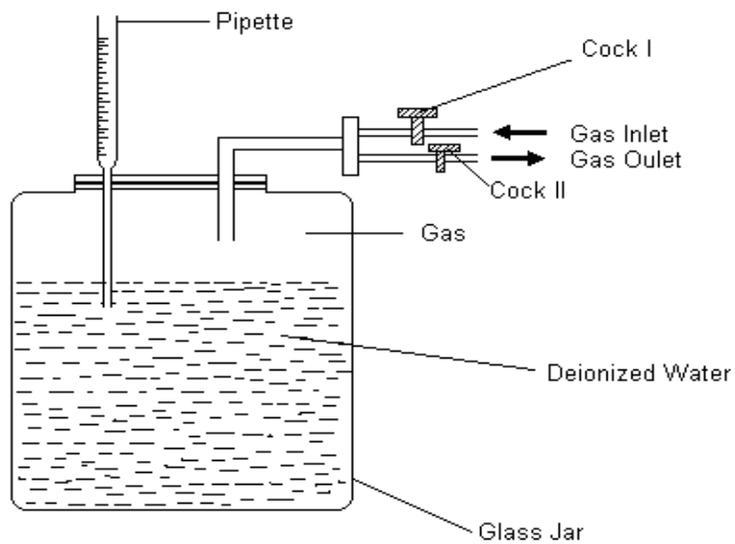

**Figure 4(b)** Concentration Measuring System.

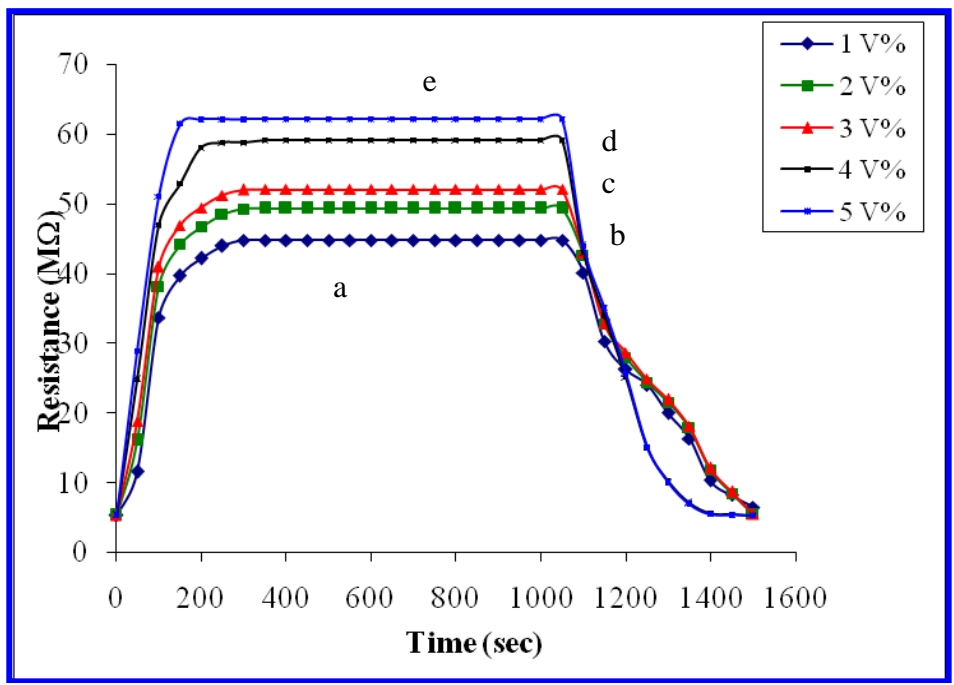

**Figure 5** Variation of resistance of $TiO_2$ with the time after exposure for different vol.% of LPG.

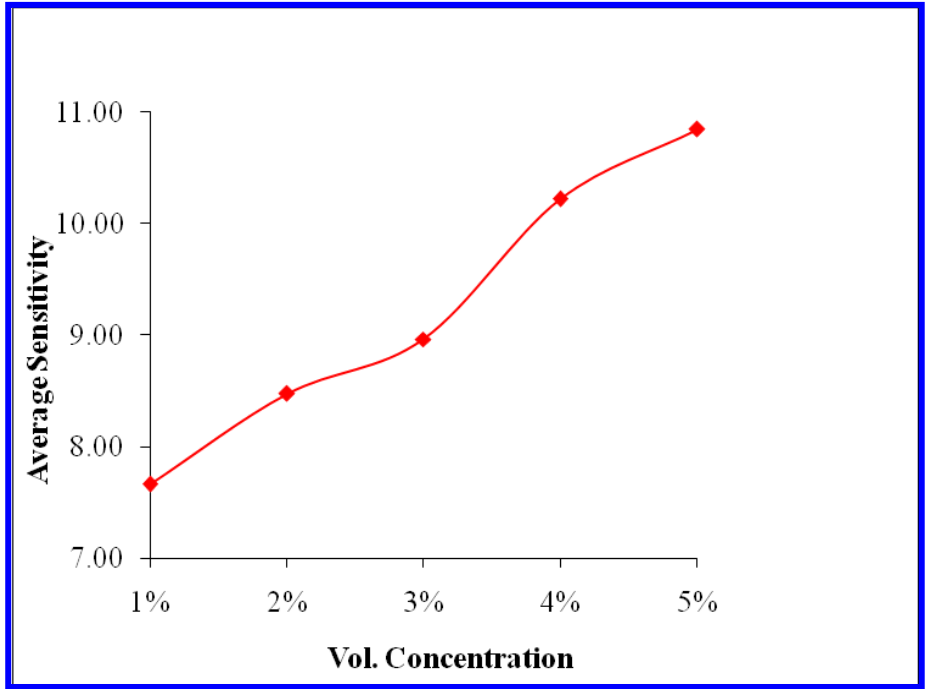

**Figure 6** Variations of average sensitivity of $TiO_2$ sensing material with concentration of LPG.

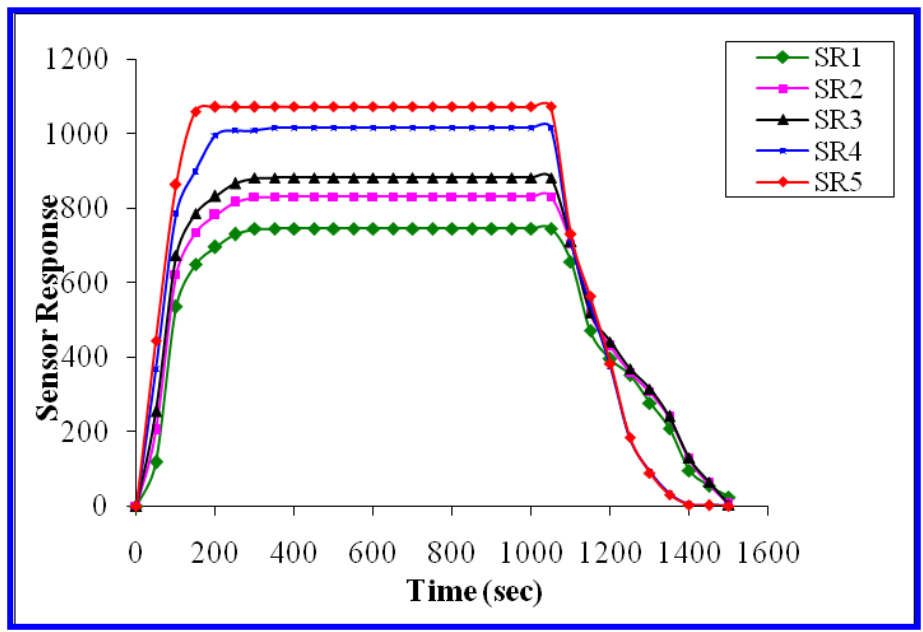

**Figure 7** Sensor Response curves of $TiO_2$ with the time after exposure for different vol.% of LPG.

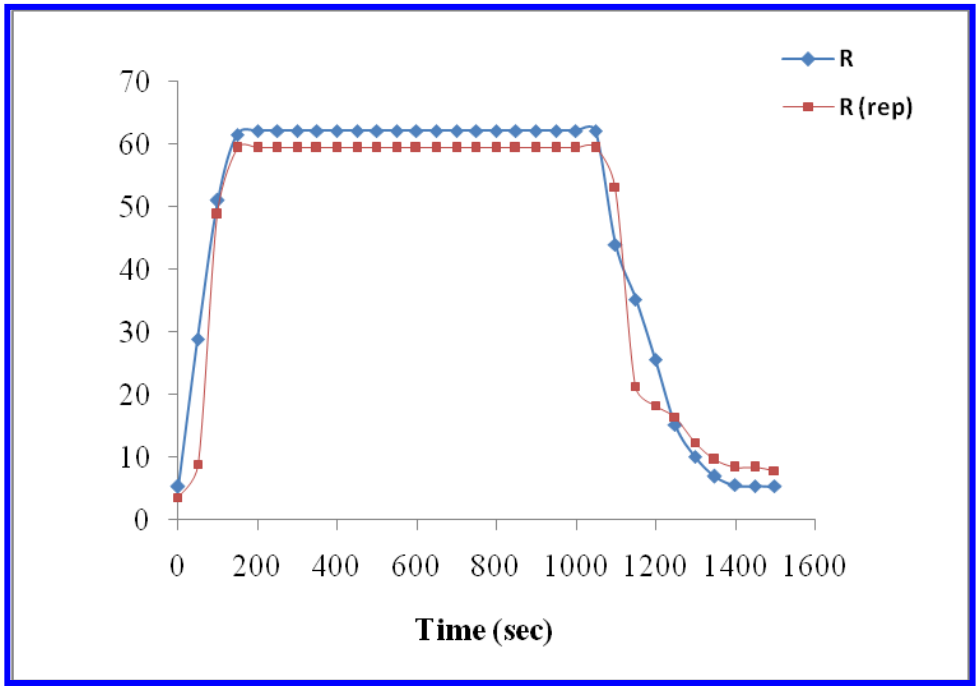

**Figure 8** Reproducibility curve for the sensing material for 5 vol. % of LPG.